\documentclass[runningheads]{llncs}
\usepackage{cite}
\usepackage{amsmath,amssymb,amsfonts}
\usepackage{algorithmic}
\usepackage{graphicx}
\usepackage{textcomp}
\usepackage{xcolor}
\usepackage{hyperref}

\usepackage[normalem]{ulem} 
\usepackage{xcolor}

\newcommand{\ins}[1]{\textcolor{blue}{\uline{#1}}} 

\usepackage{ifthen}
\usepackage{amssymb}
\newboolean{showcomments}
\setboolean{showcomments}{true} 
\ifthenelse{\boolean{showcomments}}
  {\newcommand{\nb}[2]{
    \fcolorbox{gray}{yellow}{\bfseries\sffamily\scriptsize#1}
    {\sf\small$\blacktriangleright$\textit{#2}$\blacktriangleleft$}
   }
   
  }
  {\newcommand{\nb}[2]{}
   
  }

\usepackage{booktabs}
\usepackage{array}
\usepackage{colortbl}
\usepackage{multirow}
\usepackage{longtable}

\newcolumntype{v}[1]{>{\raggedright \hspace {0pt}}p{#1}}
\newcolumntype{G}[1]{>{\columncolor{gray90}}#1}

\definecolor{Gray}{gray}{0.8}
\definecolor{gray25}{gray}{0.25}
\definecolor{gray50}{gray}{0.50}
\definecolor{gray75}{gray}{0.75}
\definecolor{gray90}{gray}{0.9}

\newcommand{\grayrow}{\rowcolor{gray90}}

\begin{document}
\title{Defining Requirements Strategies in Agile: \\A Design Science Research Study}
\titlerunning{Defining Requirements Strategies in Agile}
\authorrunning{Muhammad et al.}
%
\author{%
Amna Pir Muhammad\inst{1}
\and
Eric Knauss\inst{1}
\and
Odzaya Batsaikhan\inst{1}
\and
Nassiba El Haskouri\inst{1}
\and
Yi-Chun Lin\inst{1}
\and
Alessia Knauss\inst{2}
}
\institute{
Dept. of Computer Science and Eng.,
Chalmers $\mid$ University of Gothenburg, 
Gothenburg, Sweden
\and
Zenseact AB, Gothenburg, Sweden
}
\maketitle              
\begin{abstract}
{Research shows that many of the challenges currently encountered with agile development  are related to requirements engineering.}  
{Based on design science research, this paper investigates critical challenges that {arise} in 
agile development from an undefined requirements strategy.}
We explore potential
{ways to address these challenges} and
{synthesize the key building blocks of requirements strategies}.
Our design science research rests on a multiple case study with three 
industrial cases {in the domains of communication technology, security services, and automotive.}
We relied on a total of 20 interviews, two workshops, participant observation in two cases, and document analysis in each of the cases to understand concrete challenges and workflows. 
{In each case, we define a requirements strategy in collaboration with process managers and experienced engineers. 
From this experience, we extract guidelines for defining requirements strategies in agile {development}.}
\end{abstract}

%
%
%

\section{Introduction}
\label{sec:1-introduction}

Agile development methodologies aim to shorten the time to market and incorporate maximum changes during the sprint to meet customer needs \cite{meyer2014ugly} and have been adapted at small-scale as well as large-scale organizations \cite{larman2010practices}. With its' focus on interactions and working software over rigid processes and extensive documentation, traditional well established Requirements Engineering (RE) processes have been neglected. Research shows that many of the challenges currently encountered with agile development  are related to requirements engineering \cite{kasauli2021requirements} for example,
misunderstanding customer needs, missing high-level requirements, and difficulty
to achieve having just enough documentation.
{In this study, we} identify 
{specific RE-related} challenges 
{and related solution strategies in agile development}. 
Based on this knowledge, we derive necessary building blocks as different viewpoints that should be considered 
{when thinking strategically about RE in agile {development}}. In this, we are inspired by test strategies, which guide testing activities to achieve the quality assurance objectives \cite{ElHaskouri2021} and which mandate that each project must have its test plan document that clearly states the scope, approach, resources, and schedule for testing activities \cite{mendez2008improving}.
We argue that defining a so called \textit{requirements strategy} similar to a test strategy for RE can be critical for successful agile development.




{In this paper, we aim to establish the concept of \emph{requirements strategy for agile development} by investigating} the following research questions based on iterative design science research in three industrial case studies.

\begin{description}
\item[RQ1] Which challenges arise from an undefined requirements strategy?
\item[RQ2] How do companies aim to address these challenges?
\item[RQ3] Which potential building blocks should be considered for defining a requirements strategy?
\end{description}

{Since we particularly target agile development, we aimed to investigate requirements challenges independent from process phases or specific documents.
Instead, we took the lens of \emph{shared understanding}\cite{glinz2015shared} to investigate different RE activities (i.e., elicitation, interpretation, negotiation, documentation, general issues).
According to Fricker and Glinz, an investigation of \emph{shared understanding} may primarily target how such shared understanding is \emph{enabled} in an organization, how it is \emph{built}, and how it is \emph{assessed} for relevance \cite{glinz2015shared}}.

{Therefore, 
our contribution 
{are guidelines on} how requirements strategies should be described for agile development.
Through building three  complementary
perspectives, 
we see that the requirement strategy guidelines capture relevant information and provide a useful overview.
We suggest that a strategy defines the structure of requirements to create a shared language, define the organizational responsibilities and ownership of requirements knowledge, and then map both structure and responsibilities to the agile workflow.}



{In the next section, we provide the related work for our study.
In Section \ref{sec:3-research-method} we elaborate on our design science research method before revealing our findings in Section \ref{sec:5-findings} in order to answer our research questions.
Then, in Section \ref{sec:4-artifact}, we present our artifact - guidelines on how to define a requirements strategy for RE in agile {development}.
Finally, we discuss and conclude our paper in Section \ref{sec:7-conclusion}.}

\section{Related Work}

{Literature shows that many companies adopt
agile methods \cite{lagerberg2013impact, jorgensen2019relationships} due to its numerous benefits, for example, flexibility in
product scope which improves the success rate of products \cite{dikert2016}, in contrast to traditional development methods \cite{serrador2015does}. Furthermore, agile methods incorporate maximum change and frequent product delivery \cite{jorgensen2019relationships}, encourage
changes with low costs, and provide high quality products in short iterations \cite{meyer2014ugly}.}
{
Due to its success, agile {methodologies are} become widely popular and adopted by both small and large companies \cite{larman2010practices}.
The term \emph{large-scale
agile} refers to agile development, which includes large teams and large multi-team projects \cite{dingsoyr2014towards}. 
Dikert et al. define large-scale agile development as agile development that includes six or more teams \cite{dikert2016}.
} 

However, despite the success of agile methods, large-scale companies also still face several challenges.
Dikert et al. (2016) \cite{dikert2016} conducted a systematic literature review of empirical studies. 
The authors identified several challenges and success factors for adopting agile on a large scale.
The most mentioned challenges are change resistance, integrating non-development functions,
{difficulty to implement agile {methods}} (misunderstanding, lack of guidance), requirement engineering challenges ({e.g., }high-level requirements management largely missing in agile {methods}, the gap between long and short term planning).
Based on {a} literature review, Dumitriu et al. (2019) \cite{dumitriu2019challenges} identified 12 challenges of applying large-scale agile {methods} at the organization level.
The most cited challenge is the coordination of several agile teams.
Kasauli et el. (2021) \cite{kasauli2021requirements} identified 24 challenges through multiple case studies across seven large-scale companies. Some of the identified challenges are building long lasting customer value, managing experimental requirements, {and} documentation to complete tests and stories.
The authors conclude that strong RE approaches are needed to overcome many identified challenges.

{When it comes to RE in agile {development}, challenges that have been identified include lack of documentation, project budget, time estimation, and shared understanding of customer values \cite{elghariani2016review,inayat2015systematic,ramesh2010agile, Kasauli,alsaqaf2019quality} }
{First attempts have been made to tackle some of the challenges of RE in agile {development},} e.g., Inayat et al. and Paetsch et al \cite{inayat2015systematic,paetsch2003requirements}.
{suggest combining traditional RE with agile {methods} and encounter challenges like} how much documentation is just enough documentation \cite{hoda2010much} to have a shared understanding of customer values.

Considering that there are many challenges related to RE that can be solved through RE approaches, this paper
{proposes to use the concept of a requirements strategy as a method to define requirements engineering practices to tackle challenges related to requirements engineering in agile} {development}.


\section{Design Science Research Method}
\label{sec:3-research-method}


Our research aims to design suitable ways of defining requirements strategies for 
{organizations with agile software development}. 
Such requirements strategies should be suitable for addressing real-world needs, incorporating state-of-the-art knowledge, and ideally being empirically evaluated in practice.
Thus, we decided that design science research \cite{wieringa2009,vaishnavi,hevner} is a good fit. 

\paragraph{\textbf{Design Science Research.}}


{Our research questions are targeted towards design science research, with RQ1 focusing on the problem domain, RQ2 investigating potential solutions, and RQ3 targeting towards deriving the artifact.}
Our artifact 
{are guidelines on how to define a} requirements strategy in agile development. 
Refining on well-known challenges with RE in agile development, we needed to gain in-depth insights into those challenges related to a lack of a clear requirements strategy throughout the agile development organization (RQ1).
Throughout our cases, we discuss those challenges with respect to potential mitigation strategies (RQ2) for those challenges.
{Finally, we systematically synthesize the building blocks of requirements strategies (RQ3) from solution strategies.} 

\begin{table}[tb]
\caption{Research questions in relation to cases and research cycles}
\label{tab:method}
\centering
{\footnotesize
\begin{tabular}{l|c|ccc|c}
\toprule
& Case 1 & \multicolumn{3}{c|}{Case 2} & Case 3 \tabularnewline
& Cycle 1 & Cycle 2 & Cycle 3 & Cycle 4 & Cycle 5
\tabularnewline
\midrule
Identify challenges (RQ1) 
&$\blacksquare\blacksquare\blacksquare\blacksquare\blacksquare\blacksquare$ 
& $\blacksquare\blacksquare\blacksquare\blacksquare\blacksquare\blacksquare$ 
& $\blacksquare\blacksquare\blacksquare\blacksquare\square\square$ 
& $\blacksquare\blacksquare\square\square\square\square$ 
& $\blacksquare\blacksquare\blacksquare\square\square\square$
\tabularnewline
\grayrow Identify solutions (RQ2)
&$\blacksquare\blacksquare\blacksquare\blacksquare\square\square$ 
& $\blacksquare\blacksquare\square\square\square\square$ 
& $\blacksquare\blacksquare\blacksquare\blacksquare\blacksquare\blacksquare$ 
& $\blacksquare\blacksquare\blacksquare\blacksquare\square\square$ 
& $\blacksquare\blacksquare\blacksquare\square\square\square$
\tabularnewline
Building blocks (RQ3)
& $\blacksquare\blacksquare\square\square\square\square$ 
& $\square\square\square\square\square\square$ 
& $\blacksquare\square\square\square\square\square$ 
& $\blacksquare\blacksquare\square\square\square\square$ 
& $\blacksquare\blacksquare\blacksquare\blacksquare\blacksquare\blacksquare$ 
\tabularnewline
\midrule
Data source 
& 
11 Interviews,
& 
\multicolumn{3}{c|}{9 Interviews, }
& 
{Participant}
\tabularnewline
& document 
&  \multicolumn{3}{c|}{participant observation, }
& observation, 
\tabularnewline
& analysis & \multicolumn{3}{c|}{document analysis,} & document 
\tabularnewline
& & \multicolumn{3}{c|}{1 workshop} & analysis,
\tabularnewline
& & \multicolumn{3}{c|}{} & 1 workshop
\tabularnewline
\bottomrule
\end{tabular}}
\end{table}
Inspired by the regulative cycle \cite{wieringa2009}, the artifact (guidelines for defining a requirements strategy based on good practices from our cases) has iteratively evolved, allowing to refine the knowledge with respect to each research question.
{Table \ref{tab:method} provides an overview of our research method. 
As can be seen, we relied on three case studies over which we distribute a total of five research cycles.
The cycles differ in how much focus is given to each of our three research questions:}
%
%

\emph{\textbf{Case 1} - Exploring the problem through the lens of requirements engineering and shared understanding:}
Case 1{, an information and communication technology company, focuses} on a strategy to achieve a shared understanding about customer value throughout the development organization.
Our research aims were two-fold: understand the real world problem and conceptualize a design artifact that may address this problem. 
Within a Master's thesis \cite{Odzaya2018}, we developed an appropriate lens that combined both the concept of shared understanding (as expressed by Glinz and Fricker through enabling, building, and assessing shared understanding \cite{glinz2015shared}) and commonly used {RE} activities (such as elicitation, interpretation, negotiation{,} and documentation).
We then relied on 11 interviews
to understand customer value and its common understanding, information sharing, tools and channels for sharing, and tools and methods for documenting. 
Since 
{our first cycle focuses on the} exploration of the problem
we locally relied on the case study research method for our research with respect to Case 1 \cite{Odzaya2018}.
{As Table \ref{tab:method} shows, we complemented the interviews with document analysis to produce an overview of challenges and related solution strategies.}

\emph{\textbf{Case 2} - Refining the requirements strategy artifact iteratively:} We then followed up in \emph{Case 2}, a company producing security smart alarms and services. 
In this case{,} the focus was on a more general requirements strategy that covers both stakeholder and system requirements.
Again, through a Master's thesis \cite{ElHaskouri2021}, we investigated concrete requirements challenges of an agile team, defined a requirements strategy along the lines of the result from Case 1, and investigated in depth to what extent it could help with the challenges in practice. 
{At this point, we 
{further focused on investigating whether} there are reusable building blocks for a requirements strategy.}

\emph{\textbf{Case 3} - Applying and evaluating the artifact:} Finally, we brought our experience from the previous two cases into \emph{Case 3}, an automotive supplier, focusing on complex safety critical and software intense systems. 
Here, 
{we focus less on challenges and solution strategies, in particular, since the case company already had compiled a good overview.
Instead, }our focus 
{is to refine the artifact (guidelines for requirements strategies) by discussing, applying, and improving our understanding of the building blocks of a requirements strategy}.
At the time of the research, the first author of this paper did an internship with this automotive supplier and helped to make the requirements strategy explicit. The company did already identify some of the challenges and making a first step towards implementing 
{their solution strategies}. 
Thus, she was able to investigate the phenomenon as a participant observer, contrasting it with documents and ways of working in the practical context,
{allowing us to fine-tune our guidelines for requirements strategies to provide an overview of challenges and solution strategies for continuous process improvement}. 
\vspace{.05cm}
\paragraph{\textbf{Data collection.}}
We relied on a mix of different methods for data collection, including interviews, participant observation, document analyses, and workshops.

\textit{Interviews -} 
We relied on interviews in Case 1 and Case 2, in particular, to understand the problem (RQ1) in each specific case.
20 interviews were conducted using interview guides 
(
{details in} \cite{ElHaskouri2021,Odzaya2018}), relying on a mix of closed and open-ended questions. 
Interviews were recorded, transcribed, and then coded.
In both cases, we recruited interviewees through purposeful sampling \cite{Palinkas2015}.
We relied on convenience sampling so that interviewees were both available and knowledgeable. 
We employed diversity sampling to capture the views of multiple relevant roles and stakeholders and similarity sampling to ensure that we received multiple views for each relevant perspective for triangulation.

\textit{Participant Observation -}
{T}he fourth author was very familiar with Case 2, in which she had worked for several years before starting her Master thesis \cite{ElHaskouri2021}.
Her work included defining a testing strategy, which provided intimate knowledge about the agile ways of working in the case company, which were also helpful for understanding the requirements-related challenges, defining a requirements strategy, and conducting the evaluation in Case 2.
The first and last authors did work with
{RE} and the continuous improvement of requirements processes of Case 3.
Through this work, 
{we were able to verify that previously identified challenges in Case 3, as well as initiatives to address them, were of similar nature and matched well with our recent work on requirements strategies.}  
{Both co-authors relied on our requirements strategy work to support the ongoing initiatives on requirements processes and on integrating RE practices into agile ways of working. This allowed us to evaluate the suitability of our requirements strategy concept.}
Knowledge from these activities was collected through field notes, presentations given at the case company, and discussions with other co-authors.

\textit{Document Analysis -}
In all three cases, a subset of the authors studied the documents related to the flow of requirements (Case 1: Author 3 and 5, Case 2: Author 4, Case 3: Author 1 and 6). 
Since all three cases embraced agile ways of working, we considered that not all relevant information might be found in formal documents.
However, we ensured that documentation did match or at least not contradict our data.
We found relevant documentation of requirements, e.g., as user stories, in all three cases to match and support our other data sources.
Document analysis also allowed us to better understand the implied requirements strategy and processes.

\textit{Workshops -}
We relied on two workshops in cases 2 and 3 to evaluate the proposed requirements strategies and, by that, also our requirements strategy guidelines.
In case 2, a workshop was conducted to present the challenges identified, the proposed solution candidates, and different versions of the specific requirements strategy of each case. In case 3, a workshop was used to understand the requirements strategy that was used to address certain challenges.
Expert participants were sampled similarly as for interviews.
They were asked to bring up additional challenges that we may have missed, give feedback on the criticality of the challenges that we had found, provide their opinion about the solution candidates, and evaluate the structure, presentation, and concrete advice on the requirements strategies. 
Depending on the circumstances in each case, we recorded and transcribed, took live notes for all participants to see, or shared our notes after the workshop for validation.

\paragraph{\textbf{Data analysis.}}
In order to analyze interview transcripts, field notes from participatory observation and document analysis, and workshop notes/transcripts, we relied on typical coding approaches for qualitative research \cite{Saldana2015}. 
This allowed us to report on challenges that relate to a missing or undefined requirements strategy.
{For example, the following quote from Case 1 contributed to identifying the challenge \emph{d) lack of communication with customers}:}
\emph{``The thing that sometimes does not work as it should, is communication with some of the customer units. It heavily depends on the competence of the customer unit people.''}

In each case, we had access to an industry champion from the respective company, who helped to suggest practical solutions. 
{For example, the following quote suggested a solution for challenge above as} \emph{{c) ability to initiate on demand meetings with customer representatives}:}
\emph{{ ``The right people to nail down a requirement should be put together in the meeting to have a requirement handshake."}}
In addition, the second author was involved as an academic supervisor in all three cases, providing pointers toward relevant published knowledge.
We regularly presented and discussed our findings at the case companies, focusing on strong tracing between challenges, solution candidates, and the proposed requirements strategies.
Together with iterative refinements, this allowed us to analyze the data in depth.

\paragraph{\textbf{Threats to Validity.}}

\emph{Internal validity} aims to reveal factors that affect the relationship between variables, factors investigated, and results.
A key threat to internal validity of this study is the risk of
{misinterpretations}, particularly during the interviews and observations. 
\emph{Construct validity} defines the extent to which the investigated measures describe what the researchers analyze and what is studied according to research questions \cite{runeson2009guidelines}.  
{We mitigated threats to internal and construct validity through interacting closely with industry partners in study design and interpretation of results. We also worked iteratively and triangulated across our iterations and cycles as well as different data sources.}
\emph{External validity}
{relates} to identifying to what extent our findings can be generalized \cite{maxwell1992understanding}.
We identified  common challenges in all three case companies.
Thus, we expect 
{that in particular the structure and perspectives of our requirements strategy guidelines can be transferred to other contexts}.
\emph{Reliability} reflects to what extent other researchers can produce the same results repeating the same study methodology. 
In a qualitative study, it is always hard to achieve \emph{reliability} since one cannot argue based on statistical significance. 
We mitigate this threat by elaborating our research method in detail to support other researchers in replicating our research and in recovering from any possible differences in results. 

\section{Findings}
\label{sec:5-findings}

\subsection{RQ1: Which challenges arise from an undefined requirements
strategy?}






The left column of Table \ref{tab:challenges} depicts RE challenges identified based on our three cases that are encountered without a clear requirements strategy existing for agile development.
{The challenges are categorized in RE practices and related to Glinz and Fricker's \cite{glinz2015shared} practices of shared understanding, grouped in three categories, i.e., enable, build, and assess.
Enable practices describe what is needed to form and establish a common foundation of knowledge. Building practices aim to provide the structured knowledge that can be communicated within the team or company through explicit artifacts or by constructing a body of implicit knowledge for shared understanding. Assessing methods determine how all team members have a shared understanding of a topic or artifact. Some methods can be used for both building and assessing practices.} 
{Indices indicate in which of the cases a challenge 
was relevant.} 



\textit{a) Teams struggle to integrate RE in their agile work efficiently$^{1,2,3}$ -}
{Agile development enables organizations to respond to change.
If there is a change in code and tests, the requirements should usually be updated. Or if requirements change, then the code and tests need to be adjusted accordingly.
Teams struggle with this since requirements tools do not integrate well with agile software development work and do not support parallel changes from several teams.
Thus, it is hard to integrate RE work into the agile work effectively.}

 \textit{b) No formal event to align on customer value$^{1}$ -}
 There were no formal events to create awareness of customer value in Case 1. 
 Even when 
 the customer unit took the initiative and organized some events, there were only a few participants. 
{Such events must be better integrated in the organization and workflow.}
 

\textit{c) Insufficient customer feedback$^{1,2}$ -}
{In Case 1 and 2, developers lack customer feedback, which is crucial for agile workflows.
This can be due to a lack of formal events, or due to scale and distance to customers.
It impacts the ability of an organization to assess whether shared understanding has been reached.}
{Customer feedback should be integrated into the workflow across organizational levels and take into account the specific needs of product owners and developers.}

\textit{d) Lack of communication with customer$^{1}$ -} 
{Customer-facing units have a key role and are on the boundary between development teams and customers.
We encounter difficulties with communication in both directions: between customer-facing teams and development teams and between customer-facing teams and the customers.}
{These challenges are mainly due to a lack of systematic guidance on how such communication should take place, thus depending completely on the individual skills of those involved.}
{Companies would have to find a way to ensure good and transparent communication, for example 
by having product owners moderating direct meetings between developers and customers.}

\textit{e) Who owns customer value$^{1}$ -} 
{Requirements enter the development organization mainly through the hierarchy of product owners (PO) in Case 1
. 
However, a significant amount of requirements originate from other sources, e.g., development teams or system managers, and in those cases, it is less clear who is able to define or {who} owns the customer value.}






\textit{f) Inconsistent elicitation$^{2}$ -} 
{POs or application specialists collect requirements when needed and apply techniques}
 such as interviews. 
 {There is, however, no systematic strategy to elicitation integrated into the workflow.}

\textit{g) Lack of feedback on elicitation$^{2}$ -}
Without a systematic validation of elicitation results, misunderstandings will only surface late in the agile workflow, e.g. during acceptance testing and result in additional costs and effort.

\begin{table*}[t!]
    \centering
    \caption{Overview of Challenges in Relation to the {Solution Strategies}. Indices ($^1, ^2, ^3$) show in which case study a challenge or strategy was encountered.}
    \label{tab:challenges}
    {\footnotesize
    \begin{tabular}{v{0.1\textwidth}v{0.24\textwidth}v{0.15\textwidth}v{0.16\textwidth}|v{0.3\textwidth}}
         \toprule 
         & \multicolumn{3}{c|}{\textbf{Shared Understanding}}  \tabularnewline
         \textbf{RE} & \emph{Enable} & \emph{Build} & \emph{Assess}  & \textbf{Solution Strategy}\tabularnewline
         \midrule
         \emph{General issues} 
         & {
         a) {Teams struggle to integrate RE in their agile work efficiently}$^{1,2,3}$}
         & b) No formal event to align on customer value$^1$
         & c) Insufficient customer feedback$^{1,2}$
         & a) Tools that allow developers to take ownership of req.$^{1,2,3}$\\ b) Regular meetings with customer representat.$^{1,2}$
         \tabularnewline \grayrow
         Elicitation 
         & d) Lack of communication with customer$^1$ \\ e) Who owns customer value$^1$  
         & {
         f) Inconsistent elicitation$^2$}
         & {
         g) Lack of feedback on elicitation$^2$}
         & c) Ability to initiate on demand meetings with customer representatives$^{1,2}$
         \tabularnewline 
         Interpretation 
         &  h) {Unclear why requirement is needed$^2$}
         &
        i) Wrong assumptions about customer value$^1$ 
         & {
         j) Unclear and volatile customer needs$^2$}
         & 
         {
         d) Fast feedback cycles$^{1,2}$}
         \tabularnewline \grayrow
         Negotiation 
         & 
         {
         k) Decentralized knowledge building$^3$}
         & l) Focus on technical details$^{1,2}$ \\ {
         m) Req. open for comments$^3$}
         & {
         n) No time for stakeholder involvement$^2$} 
         & e) Req. template includes customer value \& goals$^{1,2}$\\
        {f) Define team respon- sibilities for different parts of req. and review updates regularly$^{2,3}$ }
         \tabularnewline 
         Documentation 
         & o) Customer value description lost between systems$^1$ \\ 
         {p) Lack of knowledge about writing requirements$^{1,2,3}$
        \\q) No dedicated time for requirements$^{1,2,3}$} 
         & 
         r) Too much/not enough document.$^{1,2}$ \\ 
         {
         s) Trace the requirements to all levels, (test, and code)$^3$}
         
         & {
         t) Inconsistency b/c of requirements change$^3$ } \\
         & g) Rationale must always be provided$^1$\\
         h) Just enough documentation$^{1,2}$\\
         {
         i) Plan time for requirements updates$^3$}\\
         j) {Educate and train the development teams$^{2,3}$} \\
           {{
         k) Tools need to be setup to support traceability $^3$ 
        }}
         \tabularnewline 
         \bottomrule
    \end{tabular}
    }
\vspace{-0.7cm}
\end{table*}
\textit{h) Unclear why requirement is needed$^{2}$ -}
{Due to scale, distance to customers, or because a customer value description is not available for developers (see Challenge o), application specialists and POs may lack information on why specific low-level requirements are needed.
This can result in a gap between what product owners want and how the development teams interpret their requirements.}

\textit{i) Wrong assumptions about customer value$^{1}$ -} 
Interviewees highlighted that one of the significant challenges is that people assume customer value based on their tacit knowledge, leading to the development of faulty assumptions.

\textit{
{j) Unclear and volatile customer needs}$^{2}$ -}
{Requirements change, for example when the customer changes their mind or did not have a detailed opinion in the beginning.
When assessing the interpretation of requirements, this can cause friction, since the team tries to ``hit a moving target''.
}



\textit{k) Decentralized knowledge building$^{3}$ -}
Different teams develop requirements, architecture, and also processes at the same time. 
{This decentralized way of working is needed to yield the benefits of agile work at scale, but requires some infrastructure to enable knowledge sharing and alignment.
Otherwise, conflicting decisions will be made throughout the organization.}

\textit{l) Focus on technical details$^{1,2}$ -} 
Often customer value is not explicitly described; instead, customer needs and technical solutions are more explicit. 
When we asked participants 
{in Case 1 and 2} to 
{describe} the customer value of 
{specific} requirements, they explained the technical solutions rather than customer values. 
{This finding is consistent with documentation, where often technical details are described instead of linking to a business reason for motivating the requirement.}

\textit{m) Requirements open for comments$^{3}$ -} 
In agile development, everyone who has access to the system can create issues related to requirements in the requirements management tool.
{While it is positive to include as many stakeholders as possible in discussions, without a defined process that respects the development lifecycle, this can result in an unstructured discussion and very late changes.}

\textit{n) No time for stakeholder involvement$^{2}$ -}
Getting stakeholders' feedback after interpreting the elicited requirements is challenging since stakeholders do not have time for several meetings.
\ins{}

\textit{o) Customer value description lost between systems$^{1}$ -}
{At the scale of Case 1, it is not unusual to use several different tools to manage requirements at various abstraction layers. 
Customer-facing units use one tool, in which they define stakeholder requirements and customer value.
Development teams interact with different tools, and it is the task of the POs to refine and decompose the stakeholder requirements from tool 1 into work items for the agile teams in tool 2. 
At this step, documentation about customer value is often not transferred and thus not available to the developers.}

\textit{p) Lack of knowledge about writing requirements$^{1,2,3}$ -}
{Throughout our cases, we found that those who are responsible for documenting requirements often do not have the right training.}
{In addition, we frequently saw a lack of structure and no requirements information model.}
{Thus, teams mix stakeholder and system requirements and are challenged with writing high-quality user stories, system requirements, and in particular quality requirements.}
{In particular, the quality requirements might not get documented at all and teams will work on them without making them visible on the sprint dashboard.}

\textit{q) No dedicated time for requirements$^{1,2,3}$ -}
Since 
{agile methods focus on reducing} time to market
, spending time on writing formal requirements is not considered. 
Instead, agile teams rely on verbal requirements.
{Dedicated time to work on requirements should be integrated in the agile workflow, e.g. each sprint.}


\textit{r) Too much/not enough documentation$^{1,2}$ -}
Because agile focuses on less documentation, some essential information could be missing (e.g.,
such as the ``why'' part of the requirement). Thus, in agile {development}, determining the right amount or sweet spot of documentation is challenging. 





\textit{s) Trace the requirements to all levels, (test, and code)$^{3}$ -}
Due to ISO26262 and ASPICE compliance, the automotive company needs to guarantee full traceability between all requirements levels, (tests, and code). This places a big challenge on the entire company, since most teams work on something related to requirements, tests, or code {and those artifacts evolve in parallel}.

\textit{t) Inconsistency because of requirements change$^{3}$ -} 
{Agile methods embrace change and, consequently, teams will make changes on requirements during their work.}
However, it is challenging to handle sudden change requests and opinions from different team members{, especially at scale}. 
{The consequence can be that teams inconsistently change related requirements, or that the scope is increased without central control.}
The problem is known, yet there is a lack of guidance on how to handle this in large-scale agile development
to avoid expensive rework.




    

\subsection{RQ2: How do companies aim to address these challenges?}

{The last column of Table \ref{tab:challenges} summarizes the answers to RQ2 on solution strategies associated with the challenges with each phase of RE in respective rows{, derived from interviews, literature, or workshops and confirmed by experts in each case}.}

\textit{a) Tools that allow developers to take ownership of requirements$^{1,2,3}$ -}
{In order to allow developers to take ownership of requirements}, 
we need to find requirements tooling that integrates into the mindset and the development environment of developers to provide an efficient way of manipulating requirements. 
For instance, developers work closer to the code, so the requirements tool that supports commit/git is highly encouraged.


\textit{b) Regular meetings with customer representatives$^{1,2}$ -}
The customer-facing unit should arrange regular meetings with customers. 
{These meetings should be well integrated in the agile workflow and mandatory for team members.}

\textit{c) Ability to initiate on demand meetings with customer
representatives$^{1,2}$ -}
There should be a setup to initiate meetings with customers whenever developers need feedback.
{Since access to customer representatives is a sparse and valuable resource, a strategy for such meetings should be well aligned with the organizational structure and the agile workflow.}


\textit{d) Fast feedback cycles$^{1,2}$ -}
All teams use direct communication with stakeholders and fast feedback cycles as a baseline to get the correct interpretation.
Customer insight is abstract knowledge and could be hard to write down. There is a need to arrange events where people can meet, interact, and share customer values and feedback.

\textit{e) A requirements template that includes customer value and goals$^{1,2}$ -}
To avoid challenges related to a lack of awareness of customer value, there should be 
{specific fields or tracelinks that show how each requirement adds customer value. It is important to check their usage regularly}.




\textit{f) Define team responsibilities for different parts of requirements and review updates/comments regularly$^{2,3}$ -}
{In order to yield benefits from agile workflows, RE must be integrated into the agile workflow. 
This means that agile teams need to take responsibility of maintaining requirements and to monitor changes of requirements that are potentially related. 
This allows to manage requirements updates in parallel and at scale.
However, responsibilities have to be carefully delegated and clearly assigned.}

\textit{g) Rationale must always be provided$^{1}$ -}
The rationale for the requirement should mandatorily be provided by the role/person writing the requirement. Moreover, it should effectively be passed on from tool to tool. 

\textit{h) Just enough documentation$^{1,2}$ -}
Balancing sufficient communication and documentation is crucial in agile {development}. We should not spend too much time documenting; however, it should have all the necessary information. 
{Developers need clear guidelines to achieve this balance.}


\textit{i) Plan time for requirements updates$^{3}$ -}
Teams should plan (update, change, review) the requirements in time to align with the updated scope. 
{Such a plan should consider that updating requirements in the scope of one team may imply also requirements updates in other scopes.}



\textit{j) Educate and train the development teams$^{2,3}$ -}
{If development teams should take more responsibility of requirements, they need to be trained in RE as well as in the specifics of the overall requirements processes in their organization.
A clear requirements strategy can be a good starting point to plan such training.}

\textit{k) Tools need to be setup to support traceability$^{3}$ -}
Requirements are usually represented in different forms (e.g., textual requirements, user stories) and on different levels (e.g., system level and software level). Teams could get requirements at higher level and then derive the lower level requirements (e.g., software/technical requirements).
Tracing requirements could be hard in a large complex system. 
{Tools are needed and they should be aligned with a requirements strategy for agile workflows, i.e. allow parallel work for many teams.}

\subsection{RQ3: Which potential building blocks should be considered for defining a requirements strategy?}

This section systematically develops the building blocks of a requirements strategy from our findings in all three cases.

In Case 1, the company was challenged to establish a shared understanding. 
Proposed solution strategies for specific challenges in Case 1 can be categorized as \textbf{\emph{structural}}, \textbf{\emph{organizational}}, or related to the \textbf{\emph{work  and feature flow}}. 
For example, for the challenge \emph{l) focus on
technical details)}, a related solution strategy is \emph{e) requirements template includes customer value and goals}. 
This strategy explains that, to avoid the lack of awareness about customer value, there should be specific fields related to customer value in the requirements templates. 
This solution shows that there is a need for improvement at the \textbf{\emph{structural level}}.
In contrast, \emph{b) no formal event to align on customer value} is a challenge related to stakeholders' roles and responsibilities that needs to be well integrated into the \textbf{\emph{organization}}.
The last column in Table \ref{tab:challenges} provides a solution strategy related to this challenge as \emph{b) regular meetings with customer representative}, which relates not only to the \textbf{\emph{organizational perspective}}, but also to the \textbf{\emph{work and feature flow}}. 

In Case 2, we found the same perspectives (\textbf{\emph{structural, organizational}}, as well as \textbf{\emph{work and feature flow}}) in 
{in solution strategies for their specific challenges}.
As in Case 1, the solution strategy to introduce \emph{e) requirements templates that include customer value and goals} is a \textbf{\emph{structural}} example.
In contrast, the challenge \emph{g) lack of feedback on elicitation} can lead to misunderstandings late in an agile workflow. 
The solution strategy is to establish the \emph{c) ability to initiate on-demand meetings with customer representatives}. 
Providing access to a sparse and valuable resources such as a customer representative relates to the \textbf{\emph{organizational}} perspective.
Another related solutions strategy, \emph{d) fast feedback cycles}, for the challenge \emph{j) unclear and volatile customer needs} falls into the \textbf{\emph{work and feature flow}} perspective, by arranging events where people can meet, interact, and share customer values and feedback.

After looking deep into the concrete solution strategies in Case 1 and Case 2 {(see Table \ref{tab:challenges})}, we found that many of these strategies were already successfully implemented in Case 3.
However, the company still faced some RE challenges in agile development, allowing us to check whether the same building blocks are also applicable in Case 3. 
For example, the challenge \emph{s) trace the requirements to all levels} can be addressed with the \textbf{\emph{structural}} solution strategy \emph{k) tools to set up traceability}. 
Similarly, the challenge \emph{k) decentralized knowledge building} can be addressed by the \textbf{\emph{organizational}} solution strategy \emph{define team responsibilities for different parts of requirements and review updates/comments regularly}. 
Finally, an example of a \textbf{\emph{work and feature flow}} related solutions strategy is to \emph{i) plan time for requirements updates} in agile sprints to counter the challenge of having \emph{q) no dedicated time for requirements}.

In summary, in order to address specific challenges related to enabling, building, and assessing shared understanding of requirements in agile {development}, specific solution strategies fall into three distinct categories: \textbf{\emph{structure}}, \textbf{\emph{organization}}, 
as well as \textbf{\emph{work and feature flow}}.
Thus, a requirements strategy that bundles solution strategies for a concrete case should cover all three perspectives.

\section{Artifact: Guidelines for Defining a Requirements Strategy}
\label{sec:4-artifact}


{Our artifact is a set of guidelines for defining a \emph{Requirements Strategy} as a means to define RE activities in agile development.
As a design science research study, we built this artifact in parallel to answering our research questions iteratively.
In particular, RQ3 provides empirical validation of the building blocks.
}
At the time of research, the term ``requirements strategy'' has not been widely used. 
This is in contrast to, for example, ``test strategy'', which has quite widely been accepted to describe how testing practices can be integrated in development workflows, such as in agile ways of working. 
In our work, we refer to ``requirements strategy''  as a general strategy for including RE practices in agile methods. 

\begin{table*}[b!]
    \centering
    \caption{Building Blocks of a Requirements Strategy
    }
    \label{tab:strategy-perspectives}{\footnotesize
    \begin{tabular}{v{0.14\textwidth}v{0.2\textwidth}v{0.25\textwidth}v{0.35\textwidth}}
\toprule
& \multicolumn{3}{c}{\textbf{Support for shared understanding {of requirements}}} 
\tabularnewline
\textbf{Perspective} & \textbf{Common language} & \textbf{Knowledge flow} & \textbf{Examples}
\tabularnewline
\midrule
Structural & 
Define reqts. levels &
Define structural decomp. & 
Stakeholder, System, Component Requirements
\tabularnewline 
& 
Define reqts. types & 
Define traceability demands & 
Requirements and Traceability Information Model
\tabularnewline
& 
Define templates & 
& 
User stories include customer value and goal 
\tabularnewline
\grayrow Organizational & 
Define ownership of reqts. types & 
Define roles and responsibilities & Training plan per type/role; Team responsibility sheet
\tabularnewline
Work and feature flow & 
Define lifecycle of types & 
Map structure to workflow &
Elicitation strategy, definition of done
\tabularnewline
& & 
Map organization to workflow & 
Stakeholder map, requirements review strategy
\tabularnewline
\bottomrule
    \end{tabular}}
\vspace{-0,8cm}
\end{table*}

\textbf{Definition: Requirements Strategy.}
A requirements strategy is an outline that describes the requirements engineering approach in systems or software development cycles.
The purpose of a requirements strategy is to {support decision makers} {with} a rational deduction from organizational, high-level objectives to actual requirements engineering activities to meet those objectives and to build a shared understanding about the problem space and requirements.

The creation and documentation of a requirements strategy should be done in a systematic way to ensure that all objectives are fully covered and understood by all stakeholders. 
It should also frequently be reviewed, challenged, and updated as the organization, the ways of working, and the product evolve over time. 
Furthermore, a requirements strategy should also aim at aligning different requirements stakeholders in terms of terminology, requirements types and abstraction levels, roles and responsibilities, traceability, planning of resources,~etc. 

{Therefore, our contribution is a model of how requirements
strategies should be described for agile development. 
Through 
{providing} three 
{complementary} 
{perspectives}, the proposed 
{guidelines help to} capture relevant information and provide an useful overview.}
{Our guidelines are summarized in Table} \ref{tab:strategy-perspectives}
{, including reoccurring examples and good practices abstracted from the three case studies}.
We propose that a requirements strategy should include the following building blocks: a structural perspective, an organizational perspective, and a work and feature flow perspective.
Across these perspectives, a requirement strategy aims to support a shared understanding of requirements, in particular with respect to establishing a \emph{common language} 
{(i.e., enabling perspective in Table \ref{tab:challenges})} and with respect to facilitating the exchange and \emph{flow of knowledge} (i.e., building and assessing perspective in Table \ref{tab:challenges}).
%

We suggest to start with a structural view to create a common language.
A good starting point can be the artifacts in the development lifecycle model, for example the requirements information model in the Scaled-Agile Framework SAFe \cite{knaster2017safe}, or to define templates for user stories including customer value.
%
Based on these initial definitions, refinements can be provided based on experience, e.g., after sprint reflections.

As a second step, we propose to make the organizational perspective explicit. 
Define the roles and responsibilities with respect to the definitions in the structural view.
This can, for example, be done with a one-pager that describes the responsibilities of a team. 
Also, state who owns
which part of requirements (e.g., requirements on certain subsystems) to determine specific training needs.

Finally, the work and feature flow perspective needs to be defined. 
A good starting point can be a lifecycle model for each critical type, which is then mapped to the intended workflow. 
In agile {development}, this can partially be provided by defining done criteria.
In particular, it needs to be defined when and by whom certain information must be provided. 
If requirements elicitation efforts are anticipated, guidance should be given on obtaining the information from stakeholders.
The workflow should be related to the roles and responsibilities as well as ownership. 
A stakeholder map can provide valuable information: who owns an artifact, who should be kept informed, and who needs to review it. 
An explicit review strategy can be very valuable, affecting not only the requirements quality but also keeping reviewers informed about recent changes.
\section{Discussion and Conclusion}
\label{sec:7-conclusion}

In this design science research study, we identified challenges related to agile requirements engineering 
{in three case companies}. 
{Based on these three case studies, we identified solution strategies for resolving the identified challenges and derived building blocks as substantial parts of a requirements strategy.}
For each case we {investigated} a concrete requirements strategy. The individual requirements strategies have been well received by experts in each case company.
Specifically, we recognize the need to enable, build, and assess shared understanding of requirements in agile development. 
As our experience grew, we noticed reoccurring building blocks on what should be part of such a requirements strategy.
For our design science research, we choose therefore \emph{guidelines for creating requirements strategies} as our artifact, which we develop in parallel to investigating our knowledge questions
Our results suggest that a requirements strategy should describe how requirements are structured, how work is organized, and how RE is integrated in the agile work and feature flow. 

{Buidling on previously published challenges and solution proposals for RE in agile development (e.g. \cite{alsaqaf2019quality,kasauli2021requirements}), our contribution is to enable organizations to define a holistic approach to RE that integrates with their agile development.}
Since our guidelines shall be applicable in agile development, they do not primarily relate to explicit documentation or a dedicated requirements phase within a development lifecycle, as for example custom in waterfall processes. 
Instead, we rely on the theory of shared understanding to  {embrace} RE as a knowledge management problem and give suggestions on how organizations can approach it in their agile development.
%


Ideally, such a strategy should be documented concisely and made available to all stakeholders. 
Our requirements strategy can be interpreted as an instance of situational method engineering \cite{henderson2010situational} where we focus on the context of agile system development and requirements methods in particular. 
By this, we aim to make it easier for practitioners to integrate RE in their agile workflows.
This supports its evolution through the reflection opportunities built into agile methods. 
We hope that our requirements strategy guidelines facilitate future research on how to manage knowledge related to requirements in agile development.



%
%
\bibliographystyle{splncs04}
\bibliography{2022-SEAA-Req-Strat-references}

\end{document}